\documentclass[10pt,conference]{IEEEtran}
\IEEEoverridecommandlockouts
\usepackage{cite}
\usepackage{amsmath,amssymb,amsfonts}
\usepackage{algorithmic}
\usepackage{graphicx}
\usepackage{textcomp}
\usepackage{xcolor}
\usepackage[utf8]{inputenc} 
\usepackage[T1]{fontenc}    
\usepackage{hyperref}       
\usepackage{url}            
\usepackage{booktabs}       
\usepackage{amsfonts}       
\usepackage{nicefrac}       
\usepackage{microtype}      
\usepackage{xcolor}         
\usepackage{cleveref}
\usepackage{graphicx}
\usepackage{enumitem}
\usepackage{wrapfig}
\usepackage{pifont}
\usepackage{subcaption}
\usepackage{etoc}
\usepackage{tcolorbox}
\usepackage{tabularx}
\usepackage{array}
\usepackage{listings}
\usepackage{booktabs}
\usepackage{array}
\usepackage{makecell}

\crefname{section}{Section}{Sections}
\crefformat{section}{\S#2#1#3}
\crefname{figure}{Figure}{Figures}
\crefname{listing}{Listing}{Listings}
\crefname{table}{Table}{Tables}
\crefname{algorithm}{Algorithm}{Algorithms}
\crefname{theorem}{RQ}{Capabilities}

\def\BibTeX{{\rm B\kern-.05em{\sc i\kern-.025em b}\kern-.08em
    T\kern-.1667em\lower.7ex\hbox{E}\kern-.125emX}}
\begin{document}

\title{Guiding Human Validation of LLM-Generated Code via Verifiable Literate Programming
}

\author{
\IEEEauthorblockN{
Ziqi Yuan$^{\dagger}$,
Wenhao Lu$^{\dagger}$,
Hao Wu,
Dunhong Jin,
and Chuan Wu
}
\IEEEauthorblockA{
The University of Hong Kong\\
ziqiyuanss@connect.hku.hk,
whlu@connect.hku.hk,
wuhao55@hku.hk,
dhjin@hku.hk,
cwu@cs.hku.hk\\
$^{\dagger}$Equal contribution
}
}

\maketitle

\begin{abstract}
Vibe coding democratizes software development by allowing users to generate code via natural-language (NL) interaction with large language models (LLMs). However, the code is reliable only when it faithfully implements the user's intent, which is difficult and labor-intensive for users to validate, especially for non-programmers. Existing validation methods either rely on LLM-assisted automated testing, which suffers from prompt ambiguity and model fallibility, or involve users only in partial software artifacts such as prompts and test cases, which may overlook corner cases and program details. Motivated by a bug study of LLM-generated code, we find that detailed human feedback is essential, as the failures often stem from underspecified requirements or subtle semantic deviations, and thus cannot be resolved through automated or coarse-grained checking alone.

This paper presents \emph{verifiable literate programming} (VLP), a human-in-the-loop framework designed to make the review/validation process of LLM-generated code accessible to users at all programming levels. At its core, VLP proposes unambiguous NL-based documentation as a readable intermediate layer between prompts and code.
The documentation demonstrates concrete program semantics and enables users to provide feedback on potential intent-code mismatches. 
It supports human-involved, end-to-end repair and validation via three techniques: (i) an NL-style literate language with unambiguous syntax and mostly deterministic code-to-documentation translation, (ii) LLM-based fine-grained mismatch detection that uses trace links between prompts and documentation to focus users' review effort on suspicious documentation lines, and (iii) a verification module that leverages user-validated documentation to derive API-usage checks and formal properties, which are then verified against the generated code using model checking. Our evaluation shows that VLP improves code pass@1 from 28.7\%--73.2\% to 65.4\%--93.5\% with reasonable user effort.
\end{abstract}


\section{Introduction}
\label{sec:intro}

Vibe coding democratizes software development by shifting much of programming from manual code writing to natural-language (NL) interaction with large language models (LLMs). As a result, recent reports estimate that 63\% of vibe-coding users are non-programmers~\cite{slides:2026_coding_agent_trends, website:vibe_coding_stat}. However, LLM-generated code is useful and trustworthy only when it faithfully implements the user's intended behavior. Ensuring intent-code alignment is difficult because LLMs can hallucinate during generation, while users, especially non-programmers, often face many lines of generated code without knowing whether the program actually behaves as intended. This concern is reflected in recent developer surveys, which report that LLM-generated code remains unreliable~\cite{paper:roll_dice, paper:vibe_coding, paper:intent_form} and that 96\% of developers do not fully trust the functional correctness of AI-generated code~\cite{website:sonar_ai_code_survey}. Therefore, validating LLM-generated code has become an urgent problem for vibe coding.

This need has motivated a large body of work on automated testing and verification for LLM-generated code. One line of work utilizes LLM reasoning to generate tests or formal properties from the original NL prompt, or to directly judge program correctness~\cite{paper:learn_to_verify, paper:testgen_llm, paper:rethinking_verification_for_code, paper:testeval, paper:llm_test_generation_from_requirements, paper:tdd, paper:refine4llm, paper:lemur, paper:clover, paper:vecogen, paper:verify_ansible, paper:deepassert, paper:learn_but_verify}. However, these methods suffer from the ambiguity of the original prompt and the fallibility of LLM-generated validation artifacts. Tests may miss important paths, formal properties tend to overlook key requirements, and LLM-based judges can misjudge correctness~\cite{paper:symprompt, paper:coverup, paper:panta, paper:llm-as-a-judge-for-code}. Thus, validation that depends solely on LLM reasoning cannot provide a fully reliable basis for LLM-generated code.

Recognizing these limits, recent work has begun to involve users in specific stages of the vibe coding workflow to align code with user intent. Clarification-based methods ask users to resolve ambiguous prompts before code generation~\cite{paper:clarifygpt, paper:clarify_python, paper:curiosity_by_design, paper:clarigen, paper:llm_should_ask, paper:repair_ambiguity}, while test-driven interaction methods use user feedback on generated test cases to partially formalize intent~\cite{paper:contested, paper:persist_human_feedback, paper:testeval, paper:ticoder}. These methods can expose ambiguities or concrete program failures, but they involve users only through partial software artifacts. Prompt clarification cannot catch errors introduced during code generation, and test cases cover only sampled behaviors, potentially missing subtle deviations and corner cases~\cite{paper:symprompt, paper:coverup}. Test-case-based code validation can also be hard for data-science and domain-specific programs, where outputs are often statistical, data-dependent, and difficult to judge from input/output examples alone~\cite{paper:mltesting, paper:qcb}. 

\begin{figure}[!t]
    \centering
    \includegraphics[width=\linewidth]{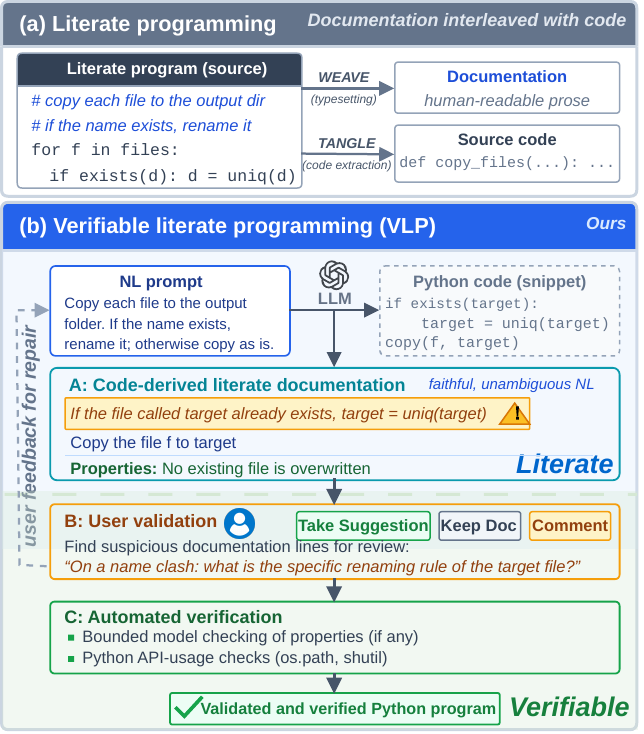}
    \caption{Conventional literate programming workflow versus verifiable literate programming (\textsc{VLP}) workflow. In \textsc{VLP}'s user validation, \textit{Take Suggestion} confirms the detected mismatch or proposed fix, \textit{Keep Doc} retains the original documentation, and \textit{Comment} indicates the user will clarify the intent.}
    \label{fig:lp_vs_vlp}
    \vspace{-5mm}
\end{figure}

Ideally, an effective validation workflow should help users find and repair more intent-code mismatches with minimal effort. To understand what failures in LLM-generated code should be exposed to users and how to effectively interact with users, we conduct a bug study on BigCodeBench-Hard~\cite{paper:bigcodebench} using DeepSeek V4 Flash~\cite{website:deepseek} and Claude Opus 4.7~\cite{website:opus}. 
We find that most failures stem from underspecified prompts or subtle semantic deviations, necessitating fine-grained user judgment beyond automated checking.
In addition, 81.7\%--83.3\% of the studied failures can be expressed as concrete intent-level behavior in NL.
These observations and literate programming~\cite{paper:literate_programming} together motivate our key idea of \emph{verifiable literate programming} (\textsc{VLP}). As illustrated in \cref{fig:lp_vs_vlp}(b), \textsc{VLP} introduces NL-based program description (called \emph{documentation}) as a faithful and readable intermediate layer between user intent and generated code. The layer is \emph{literate} because it exposes the concrete semantics of LLM-generated code in a form that users can read without programming-language knowledge. It is \emph{verifiable} because the documentation is linked to the code through (mostly) bidirectionally deterministic code-to-documentation translation. Once validated by users, it can be translated into formally verifiable properties and deterministic code aligning with user intent.

To realize this idea, \textsc{VLP} targets three connected design goals. First, documentation must be faithful and unambiguous: if the documentation is merely a lossy LLM-generated summary or comment~\cite{paper:nl_outlines, paper:ping}, subtle deviations and underspecified behaviors in the generated 
code may disappear before users can inspect them. Second, validation must be low-effort: even faithful documentation is impractical if users must review it line by line, so \textsc{VLP} should guide users to snippets most likely to require intent validation. Third, validated documentation must lead to reliable code: without translating validated or repaired documentation into reliable code, users would still need to reason about whether the code aligns with the documentation and whether programming-language-specific issues exist, such as API usage and syntax.

\textsc{VLP} realizes these goals through three components. First, it introduces an NL-style \emph{literate language} (\cref{subsec:literate_lang}) that uses syntax-directed translation to turn generated code into structured, human-readable descriptions. Meanwhile, it preserves the control flow, data flow, and assertion-based properties. Second, \textsc{VLP} provides misalignment hints through implementation-relevant traceability and taxonomy-based mismatch detection. The trace links are built between implementation-relevant sentences or formulas in the prompts and the documentation. Given these linked sentences and their surrounding context, LLMs can more accurately flag intent-documentation mismatches. Then, \textsc{VLP} highlights suspicious documentation snippets for user feedback (\cref{subsec:documentation_prompt}).  Third, it derives API checks and formal properties from the validated documentation and verifies them against the generated code with bounded model checking~\cite{paper:bounded_model_checking}. We encourage users to focus on intent-level decisions and let \textsc{VLP} automatically handle Python-specific details (\cref{subsec:documentation_code}).

We build a \textsc{VLP} prototype. Our evaluation on QuantCodeEval~\cite{paper:qcb} and BigCodeBench~\cite{paper:bigcodebench} examines whether \textsc{VLP} helps users effectively identify intent-code mismatches in LLM-generated code and improves code correctness.
The results show that \textsc{VLP} improves pass@1 from 28.7\%--73.2\% to 65.4\%--93.5\%, significantly outperforming the original generated code and three state-of-the-art (SOTA) human-in-the-loop code generation methods: ClarifyGPT~\cite{paper:clarifygpt}, TiCoder~\cite{paper:ticoder}, and PInG~\cite{paper:ping}.
Our user study further shows that \textsc{VLP} achieves higher user experience and provides better tradeoff between user validation time and code correctness.
In summary, we make the following contributions.

\begin{itemize}[leftmargin=*, noitemsep, topsep=0pt]
    \item We conduct a systematic bug study of LLM-generated code on BigCodeBench-Hard. Our study shows that most failures can be reviewed by users as intent-level misalignments through an NL representation of program semantics.
    \item We propose \emph{verifiable literate programming}, a human-in-the-loop validation framework for LLM-generated code. \textsc{VLP} first translates generated code into faithful, readable, and fine-grained literate documentation; then uses trace-link-guided LLM analysis to find likely intent-documentation mismatches for user validation and repair; and finally uses the validated documentation to derive verifiable properties and API usage checks for verifying the generated code.
    \item We evaluate \textsc{VLP} on two coding benchmarks and show that it helps users validate and repair LLM-generated code more effectively (i.e., higher pass rate) than SOTA human-in-the-loop code generation and validation methods.
\end{itemize}

\section{Background}

\subsection{Literate Programming}
\label{subsec:lp}

Literate programming~\cite{paper:literate_programming} was introduced with the philosophy that programs should be written for humans to read, not only for machines to execute. Instead of organizing a program solely around compiler requirements, literate programming asks programmers to present the program in an order and form that explain the underlying ideas to human readers. As shown in \cref{fig:lp_vs_vlp}(a), in the original literate programming system, a programmer interleaves explanatory text with code. The WEAVE processor then produces readable \textit{documentation}, while the TANGLE processor produces executable code. Although literate programming is not widely adopted today, its core idea has had broad influence. Modern computational notebooks, such as Jupyter notebooks~\cite{website:jupyter}, continue this idea by placing explanations, code, and results in the same document.

Although \textsc{VLP}, shown in \cref{fig:lp_vs_vlp}(b), is inspired by the readable documentation layer of literate programming, the two differ in several aspects. First, \textsc{VLP}'s documentation is derived from the code in a mostly deterministic and unambiguous way, rendering it more faithful to the code semantics than the free-form explanatory prose used in conventional literate programming. More importantly, literate programming uses NL primarily to support human comprehension. In contrast, \textsc{VLP} uses NL as part of a review and validation process. It assesses whether generated code implements user intent and helps identify underspecified requirements in the prompt as well as deviations in the code.


\subsection{Formal Verification Tools}
\label{bg:bmc}
Beyond improving program readability for human review, formal verification tools also play an important role in ensuring code correctness. When part of the intended behavior can be expressed as explicit formal properties, bounded model checking (BMC)~\cite{paper:model_checking, paper:bounded_model_checking} can automatically check whether a program satisfies those properties over all feasible executions up to a given bound. If a symbolic execution path reaches a violation state, the checker returns a concrete counterexample, including the input and the execution path that break the property. Otherwise, within the bounded search space explored by the checker, the property holds for all feasible executions.

Compared with unit testing, BMC provides stronger assurance. Unit tests exercise only selected concrete inputs, whereas BMC reasons over symbolic inputs and systematically explores feasible execution paths. Although BMC does not prove correctness beyond the chosen bound, it is often more automatic and lightweight than full deductive verification while providing stronger guarantees than ordinary testing.

At the same time, though formal tools can verify code with respect to a given property, they do not determine whether that property faithfully reflects the user’s intent. In particular, they cannot by themselves identify underspecification in the original prompt or detect cases where the code satisfies the written property while still subtly deviating from user intent.

\section{Bug Study and Motivation for \textsc{VLP}}
\label{sec:bug_study}

We conduct a bug study to figure out what failures in LLM-generated code should be exposed to users and how to interact with users.
Specifically, Python is chosen as the programming language for our study and system implementation because of its widespread use.
We find that failures frequently arise from underspecified requirements or subtle semantic deviations and therefore cannot be resolved by automated or coarse-grained checking alone.
Detailed NL documentation and concrete validation questions about short documentation snippets for users can expose these deviations, allowing users to confirm the intended behavior.
Our observations motivate \textsc{VLP}'s documentation-based user interface that (i) enables users to validate program semantics in NL and (ii) provides precise guidance for user feedback.

Specifically, we focus on BigCodeBench-Hard Instruct mode~\cite{paper:bigcodebench}, which contains complex instructions and diverse library usage.
We prompt Claude Opus 4.7 and DeepSeek V4 Flash (temperature=0.2).
To investigate the opportunity of NL-based documentation, we further translate each generated program into NL documentation using the approach introduced in \cref{subsec:literate_lang}.
Across 148 tasks, each with approximately five tests, we collect about 200 bugs from each model~\cite{website:opus,website:deepseek}. 
For each bug, two experienced programmers 
independently inspect the task description, test cases, error message, generated code, and corresponding documentation, and label whether the bug is reviewable in the documentation.
Disagreements are resolved through discussion.


\begin{figure}[!t]
    \centering
    \includegraphics[width=0.488\textwidth]{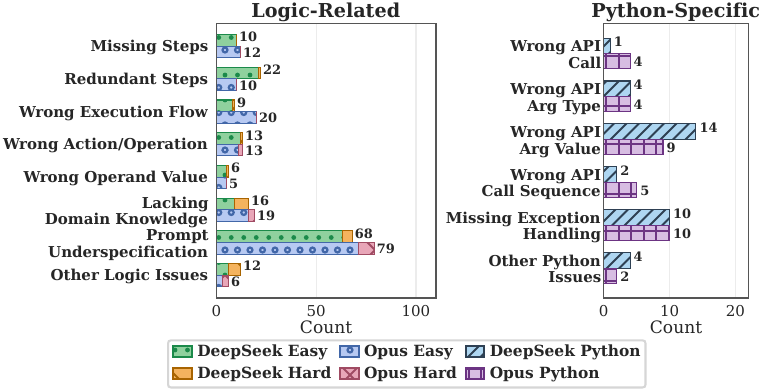}
    \caption{Bug taxonomy of code generated by DeepSeek V4 Flash and Claude Opus 4.7 using BigCodeBench-Hard~\cite{paper:bigcodebench}. 
    Note that when a prompt only omits background knowledge or conventions that are generally expected within the task domain,
    we categorize the related bug as \emph{Lacking Domain Knowledge} rather than \emph{Prompt Underspecification}.}
    \label{fig:bug_study}
    \vspace{-5mm}
\end{figure}

\subsection{Observations from the Bug Study}
\textbf{Most failures are reviewable at the intent level in NL and not solvable via automated checking}.
We first examine whether failures in the generated code can be exposed by translating code behavior into NL descriptions (\textit{documentation}).
We consider a bug documentation-reviewable if its incorrect behavior can be recognized from the documentation, without requiring users to inspect Python-specific details such as syntax, package imports, complex data types, or library usage.
As shown in \cref{fig:bug_study} (Logic-Related), most studied bugs are documentation-visible: 156 out of 191 bugs for DeepSeek and 169 out of 203 bugs for Opus.
This result suggests that documentation can expose a large fraction of failures at the intent level.
Moreover, 85.9\%--90.5\% of documentation-visible bugs are easy for users to judge because the incorrect behavior can be identified from the problematic documentation line and a few surrounding lines, without reading the full documentation.

\textbf{Failures require fine-grained review of underspecification and subtle semantic deviations}.
Generated code usually follows the overall prompt, but many reviewable failures come from underspecified or subtle details.
Prompt underspecification is the largest source, accounting for 43.6\%--46.7\% of reviewable bugs.
Across these cases, the LLM makes plausible choices according to its assumptions rather than confirmed user intent.
For example, an LLM may filter records that the user expects to keep, or silently return a fallback value instead of reporting an error.
The remaining reviewable failures often involve missing steps, redundant steps, wrong execution flow, wrong actions or operations, and wrong operand values.
Overall, we observe that many reviewable failures are subtle.
Concretely, the generated code mostly follows the prompt, but it instantiates a slightly different intent that requires fine-grained review and validation.

\textbf{Intent-level documentation cannot support reviewing Python-specific failures}.
As shown in \cref{fig:bug_study} (Python-Specific), failures that are not documentation-reviewable typically result from programming-language-level bugs that are difficult to express as user-facing intent, such as API selection, argument meanings, return-value handling, and side effects.
API calls are a typical case.
Documentation can record the callee, arguments, and return target, but judging whether they implement the intended action often requires library knowledge.
These failures are difficult to validate through documentation alone, especially for non-programmers with limited library knowledge or programming expertise.

\subsection{Design Goals}

The observations above motivate the central idea of \textsc{VLP}: using NL-based documentation as a faithful and readable intermediate layer between user intent and generated code.
For users, this layer exposes the concrete semantics of LLM-generated code in a form they can review even without programming-language knowledge. For reliable code generation, \textsc{VLP} turns confirmed user intent into program repair targets and verifiable properties.
Based on this idea, \textsc{VLP} has three design goals.

First, the \emph{code-to-documentation translation must be faithful and unambiguous} enough to demonstrate subtle deviations and underspecified behaviors in LLM-generated code. It should never use ambiguous NL sentence structures that confuse users or lose important program semantics, including control flow, data flow, and assertion-based properties.

Second, \textsc{VLP} should minimize \emph{user effort in documentation review}.
Though many failures can be exposed, users cannot be expected to review the entire documentation. Instead, an efficient framework should present users suspicious documentation snippets and provide precise surrounding context.

Finally, after multi-turn user review and repair, the validated documentation should help \emph{turn user-confirmed intent into reliable Python code}. In vibe coding, users should not be responsible for checking Python-specific details or low-level implementations of complex algorithms, as these details are hard to verify manually. Fortunately, the documentation retains assertions and concrete NL descriptions of API behaviors. Thus, \textsc{VLP} can use automated verification tools to check assertion-based properties and API usage against the generated code.

\begin{figure*}[!t]
    \centering
    \includegraphics[width=0.94\textwidth]{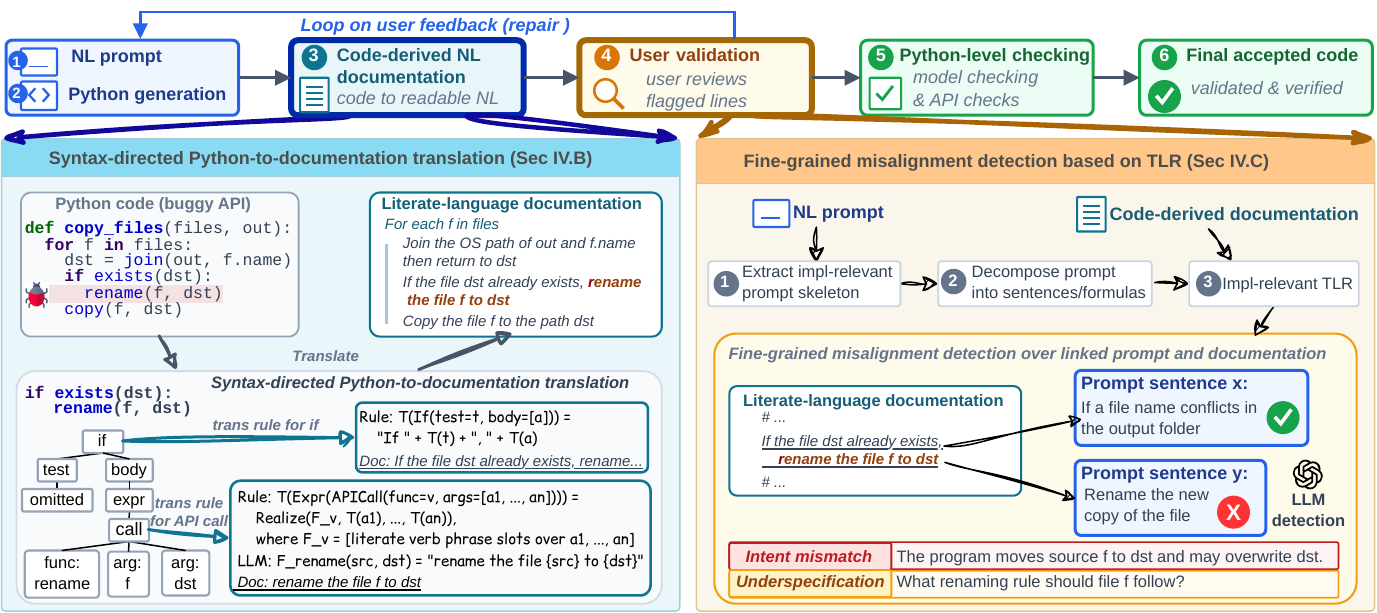}
    \caption{
    Workflow and major components of \textsc{VLP}, where the original code is buggy and the prompt is underspecified.
    }
    \label{fig:workflow}
\end{figure*}

\section{Design}
\label{sec:design}

This section first presents the overall workflow of \textsc{VLP} (\cref{subsec:workflow}) and then describes the design of three components that help users repair and validate LLM-generated code. (1) 
\textsc{VLP} introduces an NL-style language for the documentation called the \emph{literate language} (\cref{subsec:literate_lang}), which converts generated code into human-readable descriptions via syntax-directed translation. \textsc{VLP} also presents the documentation hierarchically, so users can keep the global overview while diving into local details. 
(2) 
\textsc{VLP} recovers trace links between the documentation and implementation-relevant sentences in the user prompt, allowing LLMs to identify potential mismatches more accurately by narrowing the search scope. The mismatch-detection prompt is guided by the bug taxonomy derived from our study. When a potential mismatch is detected, the LLM generates concrete validation questions about the intended behavior and presents them to users with the surrounding context (\cref{subsec:documentation_prompt}).
(3) \textsc{VLP} integrates automated tools to check API usage and algorithmic properties in the generated code against the validated documentation (\cref{subsec:documentation_code}), which reduces the burden of manually reviewing complex algorithms and Python-specific implementation details.

\subsection{Workflow of Verifiable Literate Programming}
\label{subsec:workflow}

As shown in \cref{fig:workflow}, \textsc{VLP} consists of a code generation step followed by a multi-turn review-and-repair loop until the documentation is validated by the user. Then, the code will go through API check and assertion-based property verification.

\textbf{Generation step}.
Given an NL prompt (\ding{192}), \textsc{VLP} first generates Python code, where the LLM is encouraged to generate assertion-based properties alongside the code (\ding{193}). Then, it applies Python-to-documentation translation (\ding{194}).

\textbf{Mismatch detection for guiding user review and validation}.
After the documentation is constructed, \textsc{VLP} analyzes documentation-intent alignment (\ding{195}).
It decomposes the prompt and the documentation into fine-grained alignment units and recovers trace links between them. Based on the linked pairs, \textsc{VLP} uses an LLM to detect and potential intent-documentation mismatches in a fine-grained manner.
These problematic documentation snippets are highlighted in the user interface with related validation questions and evidence from the prompt, which turns a long documentation review into a focused set of intent-level validations.
If the user revises a highlighted snippet, \textsc{VLP} patches the Python code accordingly (\ding{193}) and updates the corresponding documentation lines (\ding{194}).

\textbf{Complementary verification after user validation}.
After the documentation is validated, \textsc{VLP} checks the remaining details that are better handled by automated formal verification methods.
It uses bounded model checking (\ding{196}) for two complementary tasks: (i) Python-specific checking for library calls and related implementation details, and (ii) property checking stated as NL assertions in the documentation.
If verification fails, \textsc{VLP} feeds the failure examples back into the repair loop.
The workflow terminates once the Python code passes verification or the maximum repair count is reached.

\begin{table*}[t]
\centering
\small
\setlength{\tabcolsep}{5pt}
\renewcommand{\arraystretch}{1.12}
\sloppy
\caption{Selected literate language syntax categories, their purposes, and examples.}
\label{tab:lfl-grammar-categories}
\scalebox{0.9}{%
\begin{tabularx}{\textwidth}{>{\raggedright\arraybackslash\hspace{0pt}}p{0.17\textwidth} >{\raggedright\arraybackslash\hspace{0pt}}p{0.40\textwidth} >{\raggedright\arraybackslash\hspace{0pt}}p{0.37\textwidth}}
\toprule
\textbf{Categories} & \textbf{Purpose} & \textbf{Example} \\
\midrule
Declarations and type signatures & Introduce functions, custom types, fields, inputs, and outputs in a form close to English. & \textit{There is a calculate\_score function, whose input is a decimal named raw\_val, whose output is a decimal} \\
\midrule
Core expressions and data access & Express arithmetic, Boolean logic, etc. Prefer keeping symbolic representations as users are familiar.  & \textit{today\_score = basic\_score + correct\_answers[date, person\_id] / 2} \\
\midrule
Programming-style function calls & Support precise function invocation inside expressions, including recursion and method calls. & \textit{recursive\_calc(n - 1) + recursive\_calc(n - 2) + 1} \\
\midrule
Natural-language-style operations & Describe workflow steps and API-like operations as verb phrases, optionally with parameters and returns. & \textit{Normalize input\_values with scale = 2.0 then return to normalized} \\
\midrule
Conditionals and predicate formulas & Encode branching logic using either comparison expressions or restricted natural-language predicates. & \textit{If (volatility\_index exceeds threshold by 10 percent) and (min\_score > 50), is\_valid = true} \\
\midrule
Loops and iteration & Express both conventional loops and natural-language iteration over collections. & \textit{For each document in document\_collection, extract keywords from document and append to kw\_list} \\
\midrule
Properties assertions (Hoare logic style~\cite{paper:hoare_logic}) & State preconditions and postconditions for verification inside formally declared functions. & \textit{At this point, assume that unsorted\_score\_list.length > 0} \\
\bottomrule
\end{tabularx}
}
\fussy
\vspace{-4mm}
\end{table*}

\subsection{Literate Language Design and Documentation Construction}
\label{subsec:literate_lang}
To translate generated code into literate-language documentation, first, we design grammar and conversion rules to support (mostly) deterministic translation for validation-relevant code snippets. Second, we encourage the code-generation LLM to summarize complex algorithm implementations by extracting their properties, so users can simply review high-level behavior and automated checks can subsequently verify low-level details. Third, we organize the documentation into a hierarchical structure when showing it to users, preserving a global overview while enabling rapid comprehension of the local context.

At the language level, we define the literate language with an \textit{NL-style LALR(1) grammar}, a parser-friendly grammar with an unambiguous parse structure. This keeps the documentation readable to users while remaining precise enough to review. In this paper, we use English in the literate language and study 200 NL-program pairs, together with prior NL-programming literature~\cite{paper:nl_comp, paper:ibm_nlp}, to analyze how people express programming-language semantics in English. Nevertheless, \textsc{VLP} can be extended to other natural languages by building dedicated LALR(1) lexers and parsers.


\textbf{Syntax-directed python-to-documentation translation}.
For major Python syntax rules, we propose corresponding translation rules in the literate-language grammar.
As shown in \cref{tab:lfl-grammar-categories}, these rules follow Python's key syntactic categories, including declarations, expressions, function calls, conditionals, loops, and assertions.
During runtime, after an LLM-generated Python program is parsed into a syntax tree, \textsc{VLP} performs a syntax-directed translation. It traverses the tree and emits literate-language snippets following the corresponding translation rules.
During translation, several hierarchical Python syntax-tree nodes often need to be linearized into one literate-language sentence.
We therefore carefully design the literate-language syntax so that such sentences read naturally while still preserving the nested grammatical structure needed for LALR(1) parsing.
Specifically, step \ding{194} in \cref{fig:workflow} illustrates how a complex Python statement is translated by fixed rules into a literate-language sentence step by step on the tree.

Ideally, the translation would be fully deterministic even though Python syntax supports flexible parts such as variable names and expressions.
In most cases, this is possible because the translation rules directly carry Python variable names, expressions, and conditions into the corresponding documentation without decimating readability.
However, some flexible parts cannot be handled by syntax-directed translation alone.
Simply copying these Python snippets would make the documentation hard to read, yet their forms are too flexible for us to exhaustively define fixed rules that translate them into understandable literate-language descriptions.
API calls are a typical case.
The converter can parse the callee, arguments, and return target, but the API name often needs to be explained as a clearer action, such as creating a dataframe or computing a distance matrix.
Likewise, lambda expressions and higher-order expressions may encode sorting keys, filtering rules, or data transformations.
A fixed translation rule would either stay too close to raw Python or miss the behavior users need to validate.
\textsc{VLP} therefore statically analyzes syntax trees and marks these snippets as \textit{python\_level} blocks.
For each \textit{python\_level} block, the LLM generates a short verb phrase using its pretrained knowledge and available docstrings.
Meanwhile, for determinism, translation rules still fix the surrounding control flows, input arguments, and return assignment.

\textbf{Property support in documentation}.
Reviewing every implementation step of a complex algorithm can make user validation difficult. Therefore, \textsc{VLP} encourages the Python-generation LLM to abstract algorithms into compact properties. We use assertion-based Hoare-logic-style properties: a precondition specifies what must hold before function execution, a postcondition specifies what must hold after the function returns, and an invariant specifies what must remain true during execution.
For example, a sorting function can be documented by requiring the output to be ordered and to contain the same elements as the input, rather than describing each implementation step. Such properties serve as a shared interface between users and verification tools. In \textsc{VLP}, users validate whether the properties reflect their intent, while automated verifiers leverage bounded model checking to verify the code.




\textbf{Hierarchical documentation display for focused review}.
After translating Python code into documentation (step \ding{194} in \cref{fig:workflow}) and generating validation questions (step \ding{195}), \textsc{VLP} presents them in a hierarchical view as the user interface.
The goal is to avoid making users read the full documentation while still preserving the surrounding program context needed to judge a potential documentation-intent mismatch.
Specifically, documentation lines are grouped into collapsible blocks using function calls and classes as boundaries.
For recursive function calls and class references, the same function or class is expanded to depth at most two.
When a validation question generated in \cref{subsec:documentation_prompt} highlights a suspicious documentation snippet, the user interface proactively opens the relevant block and nearby context.
Thus, users only have to review and validate a small portion of the documentation.

\subsection{Fine-Grained Intent-Documentation Mismatch Detection}
\label{subsec:documentation_prompt}

After \textsc{VLP} constructs the documentation, users will validate whether its behavior matches their intent.
A full documentation view is insufficient for complex programming tasks, as users do not know which parts of the long documentation deserve more attention.
To narrow the scope, as illustrated in step \ding{195} of \cref{fig:workflow}, \textsc{VLP} conducts fine-grained misalignment detection based on implementation-relevant trace links between documentation and prompts, and then presents found mismatches and corresponding validation questions via the user interface.

\textbf{Implementation-relevant traceability link recovery}.
\textsc{VLP} detects mismatches by comparing linked pairs of prompt-side and documentation-side link units.
Each unit covers a single action, a condition, a function signature, or a formula, including constants and literals.
For example, a sentence in the prompt describing how invalid inputs should be handled should be linked to the code snippets that implements the check and corner-case handling. 
Such fine-grained links have two benefits.
First, they allow the LLM to focus on a specific bug type and a small snippet in each detection step, following the general principle of decomposing complex tasks into focused subproblems~\cite{paper:thought,paper:ltm_prompt}.
Second, the narrow scope makes the detected mismatches easier for users to review.


\newcommand{\lcell}[1]{\begin{tabular}[t]{@{}l@{}}#1\end{tabular}}
\begin{table*}[t]
\centering
\footnotesize
\setlength{\tabcolsep}{3pt}
\renewcommand{\arraystretch}{0.9}
\caption{Fine-grained mismatch detection in \textsc{VLP}. A unit may fall into multiple categories; for example, a line involving both formula logic and API calls is checked under both \emph{linked pair with API use} and \emph{other linked pair}.}
\label{tab:mismatch_detection}

\begin{tabular}{@{}
>{\raggedright\arraybackslash}p{0.12\textwidth}
>{\raggedright\arraybackslash}p{0.19\textwidth}
>{\raggedright\arraybackslash}p{0.47\textwidth}
>{\raggedright\arraybackslash}p{0.18\textwidth}
@{}}
\toprule
\textbf{Scope} &
\textbf{Mismatch type} &
\textbf{LLM check criterion} &
\textbf{Validation question} \\
\midrule

\lcell{Unlinked\\prompt unit} &
\emph{Missing steps} &
A prompt-side unit has no linked documentation unit. The LLM checks whether the behavior is truly required but missing. &
Whether the prompt unit should be implemented. \\

\midrule
\lcell{Unlinked\\doc unit} &
\emph{Redundant steps} &
A documentation-side unit has no prompt support. The LLM checks whether it is truly redundant behavior or a reasonable enhancement. &
Whether the extra program behavior is intended. \\

\midrule
\lcell{Unlinked or\\ambiguous unit} &
\lcell{\emph{Prompt underspecification};\\\emph{lacking domain knowledge}} &
The prompt does not fully determine the implementation choice, or the behavior depends on domain-specific conventions. &
Corresponding clarification questions. \\

\midrule
\lcell{Linked pair\\with API use} &
\emph{Wrong action/operation} &
For \emph{python\_level} blocks that invoke APIs, \textsc{VLP} retrieves API docstrings and semantically similar alternatives. The LLM checks whether the current API behavior matches the intended operation. &
Which operation or documented API behavior is intended. \\

\midrule
\lcell{Linked pair\\with API use} &
\emph{Wrong operand/value} &
The API-level action is correct, but parameters, operands, exception behavior, or boundary cases differ from the prompt. &
The intended details or edge-case behavior. \\

\midrule
Other linked pair &
\lcell{\emph{Wrong action/operation};\\\emph{Wrong operand/value}} &
The documentation simplifies or misinterprets the required operation. For example, the prompt asks to combine $A$ and $B$ without specifying the formula, while the documentation instantiates it as $(A+B)/2$. &
The intended operation, operand, or property. \\

\midrule
Function-level flow &
\emph{Wrong execution flow} &
The prompt-side control-flow sketch and documentation-side execution flow disagree on conditions, loops, or ordering. &
When or in what order the behavior should occur. \\

\bottomrule
\end{tabular}
\vspace{-3mm}
\end{table*}

We regard this as a traceability-link recovery (TLR) problem~\cite{paper:tlr} with an implementation-relevance filter, which we refer to as implementation-relevant TLR. Our design is motivated by the observation that complex coding prompts often contain implementation-irrelevant noise, such as explanatory sentences and background information.
Thus, instead of treating all prompt sentences as link candidates, \textsc{VLP} identifies the behaviors and constraints that a correct implementation must realize and builds link units around them.
Specifically, the prompt side and the documentation side are selected and decomposed differently.
For documentation, where program structure is available, \textsc{VLP} uses syntax trees to identify candidate link units such as verb-phrase statements, conditions, properties, formulas, and basic blocks.
These units expose the concrete actions and conditions expressed by the implementation.

On the prompt side, \textsc{VLP} uses an LLM to decompose the prompt into hierarchical link units. Each unit exposes one implementable behavior or constraint, while preserving its text from the original prompt. When a sentence specifies multiple behaviors, \textsc{VLP} further splits it into smaller clauses.
Unlike flat sentence-level linking, this hierarchy lets \textsc{VLP} keep each link unit concrete while retaining the logical context needed to interpret it. This design serves two purposes. First, it helps \textsc{VLP} handle long prompts without collapsing fine-grained requirements into coarse summaries. Instead of asking the LLM to produce a single global summary, \textsc{VLP} first identifies coarse implementable behavior regions and then recursively decomposes each region into more detailed child units. The recursion stops once the relevant text span reaches a manageable size for the LLM to output extracted sentences from the original prompt rather than free-form summaries. Second, the hierarchy preserves logical relations among prompt-side units. When one unit specifies the condition under which another behavior should occur, such as an error condition followed by a required message, the behavior unit is recorded as a child of the condition unit. The hierarchy retains the condition/context in which each behavior should be implemented.


Given the prompt-side and documentation-side candidate units, \textsc{VLP} retrieves the top-$k$ documentation candidates for each prompt unit based on semantic similarity, where $k=3$ in our evaluation. Following recent LLM-assisted TLR work~\cite{paper:lissa}, an LLM-based judge then confirms or rejects each candidate link using the two units and their surrounding context. Using LLM-assisted TLR allows the recovered links to reflect relevant program behavior rather than superficial textual similarity.

\textbf{Fine-grained misalignment detection}.
After recovering trace links, \textsc{VLP} uses an LLM to check both linked prompt-documentation pairs and unlinked units for potential mismatches. Guided by the logic-related bug taxonomy in \cref{fig:bug_study}, \textsc{VLP} considers seven mismatch types. \Cref{tab:mismatch_detection} summarizes the fine-grained detection criteria for each type.

We additionally introduce the way to detecting underspecification  for \emph{linked pairs with API use}, which is not illustrated by the table. For a linked pair whose documentation-side unit invokes an API-based \texttt{python\_level} block, \textsc{VLP} retrieves the API docstring and top three semantically similar alternatives from an incrementally extended API knowledge base. If the retrieved APIs suggest multiple plausible behaviors or edge-case handling, but the prompt-side unit does not specify which one is intended, \textsc{VLP} treats the case as API-level \emph{prompt underspecification} and asks the user for clarification.

\subsection{Automated Verification After User Validation}
\label{subsec:documentation_code}

Documentation enables users to validate intent-level behavior, but low-level concerns such as Python-specific API requirements and the correctness of algorithms are better handled automatically. To this end, \textsc{VLP} integrates two complementary verifiers: an API verifier for Python-specific API usage, and a bounded model checker for postconditions and invariants stated in the validated documentation.

\textbf{Python-specific API verifier}.
User validation helps ensure that the selected APIs match the intended functionality, but users, especially non-programmers, may still be unfamiliar with Python-specific API requirements, such as stateful APIs, argument and return-value constraints, and initialization and cleanup. Thus, \textsc{VLP} checks API-level requirements automatically with an LLM in the code repair process.

\textbf{Bounded model checking for properties}.
During constrained Python generation, \textsc{VLP} emits verifiable properties, including preconditions, postconditions, and invariants, as executable assertions. After users validate these properties in the documentation, \textsc{VLP} checks the corresponding functions with bounded model checking. Concretely, we use CrossHair~\cite{paper:python_crosshair} to symbolically explore program executions and search for inputs that violate the validated assertions. When CrossHair finds a counterexample, \textsc{VLP} feeds the failing input and violated assertion back to the LLM for subsequent program repair.


\section{Evaluation}
\label{sec:evaluation}

Our evaluation answers the following research questions:
(i) Does \textsc{VLP} generate more correct programs than LLM-only generation and prior human-in-the-loop code generation frameworks (\cref{subsec:passk})?
(ii) How accurately does \textsc{VLP} localize intent-documentation mismatches (\cref{subsec:precision})?
(iii) How much does \textsc{VLP} affect user satisfaction during repair and validation (\cref{subsec:precision})?
(iv) What additional cost does \textsc{VLP} introduce (\cref{subsec:cost-eff})?
(v) How much do the API knowledge base and implementation-relevant TLR contribute to \textsc{VLP}'s effectiveness (\cref{subsec:ablation})?

\subsection{Evaluation Setup}
\label{subsec:evaluation_setup}

\textbf{Implementation and evaluated LLMs}.
We implement \textsc{VLP} using LangGraph~\cite{website:langchain} and NLTK~\cite{book:nltk}.
We evaluate it with GPT-5.4~\cite{paper:gpt5} and DeepSeek V4 Flash~\cite{website:deepseek} by calling the official APIs with default decoding parameters.

\textbf{Benchmark selection.}
We evaluate \textsc{VLP} on two complementary LLM-for-code benchmarks.
(1) We use BigCodeBench-Instruct~\cite{paper:bigcodebench} (BCB) to evaluate LLM-generated code on tasks with diverse API calls and complex instructions.
(2) We use QuantCodeEval~\cite{paper:qcb} (QCE), a finance paper-to-code benchmark, to evaluate domain-specific and extremely complex code generation tasks.


\textbf{Metrics}.
We evaluate \textsc{VLP} from four perspectives: end-to-end code correctness, guidance quality, user experience, and cost efficiency.
For \emph{end-to-end code correctness}, we report pass@1~\cite{paper:humaneval} using benchmark-provided tests and property checkers.
Pass@1 measures whether a single generated implementation passes all required checks.
For \emph{guidance quality}, we measure the precision of \textsc{VLP}'s mismatch detection (recall is already represented by pass@1), defined as the fraction of suspicious documentation snippets reported by \textsc{VLP} that correspond to ground-truth intent-documentation mismatches.
For \emph{user experience}, we demonstrate code review and validation time and users' self-reported experience in the user study.
Finally, for \emph{cost efficiency}, we report the average token usage and dollar cost per programming task.

\textbf{Baselines}.
We compare \textsc{VLP} against four baselines.
\emph{Default Code} directly prompts the LLM to generate code without user review, clarification, or validation.
We also compare against three open-source SOTA human-in-the-loop code generation frameworks, covering pre-generation clarification, comment-based validation, and test-driven interaction.
\emph{ClarifyGPT}~\cite{paper:clarifygpt} is a strong pre-generation intent clarification baseline.
\emph{PInG}~\cite{paper:ping} is a comment-based human-in-the-loop framework that interleaves code generation, comments, and user feedback through editable comments, and is reported to outperform Copilot.
We include it to test whether ordinary editable comments are sufficient, or whether \textsc{VLP}'s documentation, guidance, and verifiers provide additional benefits.
\emph{TiCoder}~\cite{paper:ticoder} is a test-based intent clarification framework that uses generated tests and user feedback on the tests to infer user intent and refine code.
We include it to compare documentation-based validation against test-driven interaction.

\begin{table}[!t]
\centering
\scriptsize
\setlength{\tabcolsep}{2.0pt}
\renewcommand{\arraystretch}{1.08}
\caption{Tool and task assignment for users. Clari: ClarifyGPT; TiCo: TiCoder. P/L denote participants with professional and low programming experience, respectively.}
\label{tab:real-user-assignment}
\scalebox{0.9}{%
\begin{tabular}{lcccccccc}
\toprule
& \textbf{{BCB-21}} 
& \textbf{{BCB-254}} 
& \textbf{{BCB-344}} 
& \textbf{{BCB-389}} 
& \textbf{{BCB-493}} 
& \textbf{{BCB-597}} 
& \textbf{QCE-T01} 
& \textbf{QCE-T24} \\
\midrule
P1 & \textsc{VLP} & TiCo & PInG & Clari & \textsc{VLP} & TiCo & PInG & Clari \\
L1 & \textsc{VLP} & TiCo & PInG & Clari & \textsc{VLP} & TiCo & PInG & Clari \\
P2 & Clari & \textsc{VLP} & TiCo & PInG & Clari & \textsc{VLP} & TiCo & PInG \\
L2 & Clari & \textsc{VLP} & TiCo & PInG & Clari & \textsc{VLP} & TiCo & PInG \\
P3 & PInG & Clari & \textsc{VLP} & TiCo & PInG & Clari & \textsc{VLP} & TiCo \\
L3 & PInG & Clari & \textsc{VLP} & TiCo & PInG & Clari & \textsc{VLP} & TiCo \\
P4 & TiCo & PInG & Clari & \textsc{VLP} & TiCo & PInG & Clari & \textsc{VLP} \\
L4 & TiCo & PInG & Clari & \textsc{VLP} & TiCo & PInG & Clari & \textsc{VLP} \\
\bottomrule
\end{tabular}
}
\end{table}

\begin{figure}[!t]
    \centering
    \includegraphics[width=\linewidth]{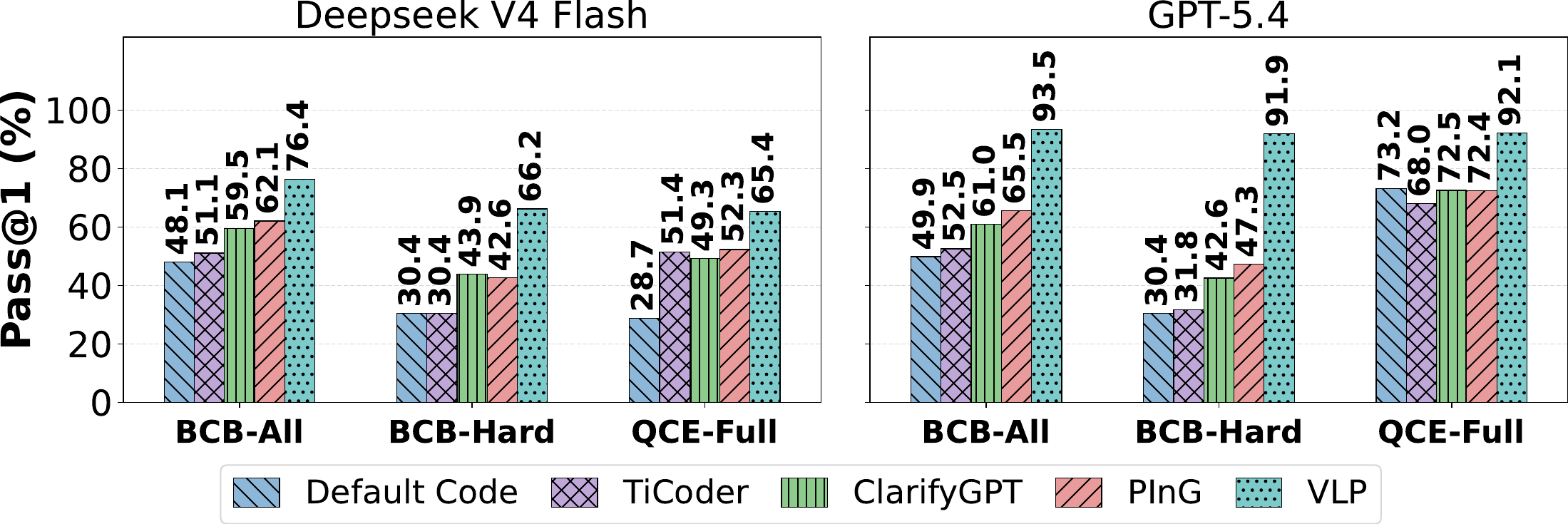}
    \caption{Pass@1 on BCB and QCE with user simulation.}
    \label{fig:passatone}
    \vspace{-5mm}
\end{figure}

\begin{figure}[!t]
    \centering
    \includegraphics[width=\linewidth]{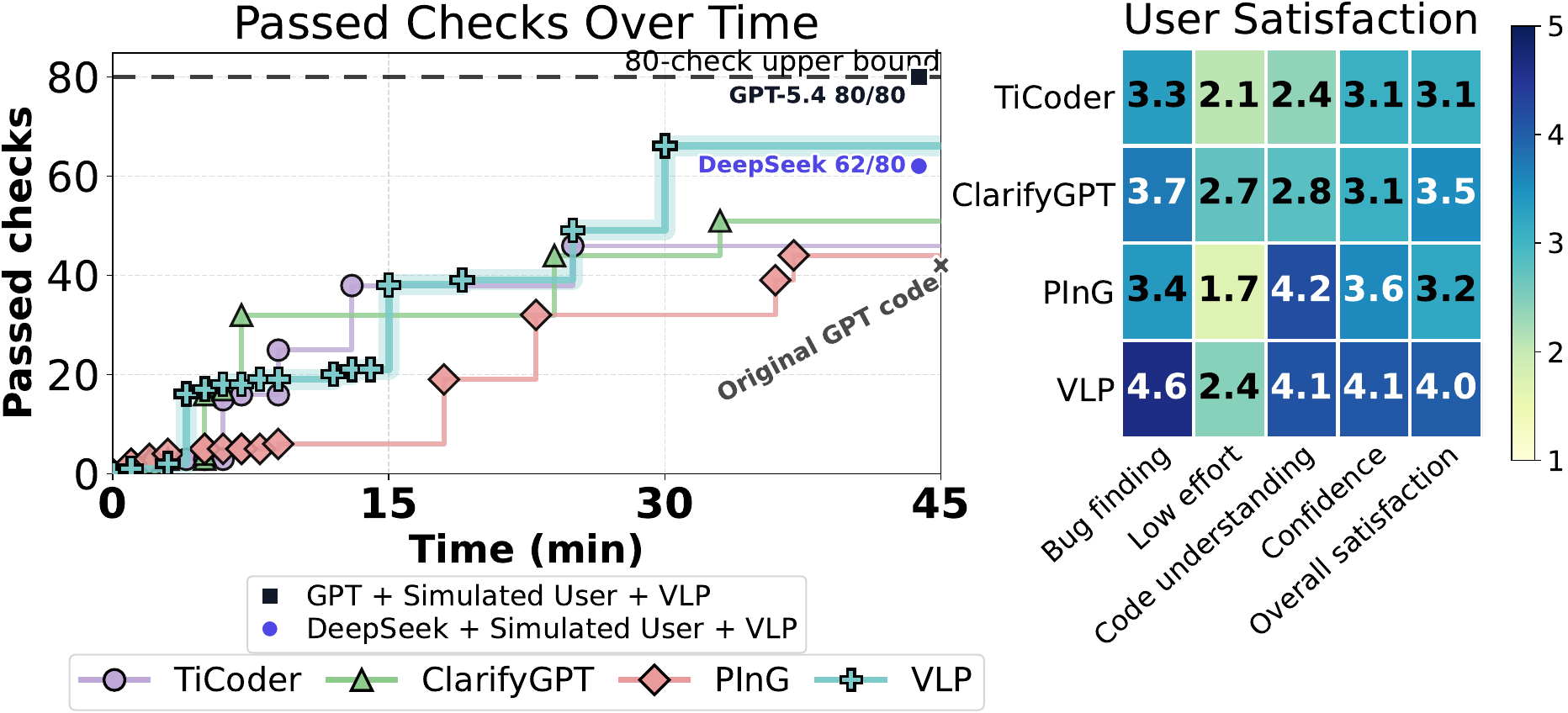}
    \caption{Real user study showing pass@1 over user time (left) and users’ satisfaction (right). The final pass rate of \textsc{VLP} falls between those in the simulated user settings of DeepSeek and GPT.}
    \label{fig:user_study}
    \vspace{-5mm}
\end{figure}

\begin{table*}[!t]
\centering
\small
\setlength{\tabcolsep}{3.2pt}
\renewcommand{\arraystretch}{1.12}
\caption{Method-specific inputs (in addition to the oracle context) and outputs for LLM-based user simulation.}
\label{tab:user-simulation-setup}
\scalebox{0.87}{%
\begin{tabularx}{1.08\textwidth}{
>{\raggedright\arraybackslash}p{0.09\textwidth}
>{\hsize=.70\hsize\raggedright\arraybackslash}X
>{\hsize=1.30\hsize\raggedright\arraybackslash}X}
\toprule
\textbf{Method} & \textbf{Simulator input following the papers} & \textbf{Allowed simulator output} \\
\midrule
\textsc{VLP}
& Validation questions and highlighted documentation (rather than the code).
& Answer only the validation questions. The simulator may fix the highlighted statements according to the validation hints, but cannot touch unrelated documentation parts. \\
\midrule
ClarifyGPT
& Clarification questions generated from the ambiguous requirements~\cite{paper:clarifygpt}.
& Answer only the asked clarification according to the intended behavior. The simulator cannot provide unsolicited bug fixes or compare the full code with the reference. \\
\midrule
PInG
& NL comments and the binary PASS/FAIL execution results~\cite{paper:ping}.
& Review only the shown comments. The simulator may confirm or edit each comment line, but cannot perform code-level review. \\
\midrule
TiCoder
& Five candidate tests generated by TiCoder with five candidate code implementations~\cite{paper:ticoder}.
& Return the expected pass/fail outputs of the golden reference code given the input test cases, which follows TiCoder's open-source Github repository. \\
\bottomrule
\end{tabularx}
}
\vspace{-2mm}
\end{table*}


\subsection{Overall Pass Rate}
\label{subsec:passk}
Since our evaluation involves 30 finance papers and 1,140 general programming tasks, conducting the entire evaluation with real users would be prohibitively costly. We therefore combine real user feedback with an LLM-based user simulator for large-scale evaluation, following prior interactive code-generation work~\cite{paper:clarifygpt,paper:ticoder}. The consistency analysis between simulated and real users is in \cref{subsec:user consistency}.

\textbf{LLM-based user simulation}.
As shown in \cref{tab:user-simulation-setup}, the simulator is provided with oracle context which represents the full vibe coding user intent. The oracle context includes the benchmark prompt and golden reference implementation, so that the user simulator's feedback is grounded in the intended behavior.
This oracle context is available only to the simulator, not to the evaluated framework during code generation or repair.
The simulator can only respond to questions posed by \textsc{VLP}, ClarifyGPT, and TiCoder, and update PInG’s comments via their original interaction interface for fair comparison.
Moreover, it cannot reveal unrelated code defects proactively.

\textbf{Real user study}.
We conduct a real-user study on a selected subset to compare \textsc{VLP} with prior human-in-the-loop methods.
We recruit 10 participants, reserving 2 for pilot testing to refine task selection, user interfaces, and instructions, and using the remaining 8 in the main study.
We sample 8 tasks from the benchmarks, including 2 QCE and 6 BCB tasks that are unsolved by the initial code. Before the study, participants provided informed consent and completed a pre-study questionnaire on their background.
For each human-in-the-loop method, we implement a user interface following the paper.
The model used in the real-user study is GPT-5.4.

We assign tasks and methods to participants according to three criteria, as shown in \cref{tab:real-user-assignment}.
First, each selected task is evaluated with all four human-in-the-loop methods, so method comparison is not confounded by task difficulty.
Second, each participant sees only one method for the same task, avoiding carryover effects from answering the same task multiple times.
Third, each method on each task is assigned to one programmer and one non-programmer participant, reducing confounding from participant expertise.
Each participant will respond through the dedicated user interface.


\begin{table}[!t]
\centering
\small
\setlength{\tabcolsep}{2.2pt}
\renewcommand{\arraystretch}{1.12}
\caption{Simulator-human consistency. H denotes human-human consistency; S denotes simulator-human consistency.}
\label{tab:simulator-validation}
\scalebox{0.8}{%
\begin{tabular}{lcccc}
\toprule
\textbf{Method}
& \textbf{H-correctness}
& \textbf{S-correctness}
& \textbf{H-scope}
& \textbf{S-scope} \\
\midrule
\textsc{VLP}
& 0.72
& 0.69
& 0.87
& 0.81 \\
ClarifyGPT
& 0.64
& 0.80
& 0.96
& 0.93 \\
PInG
& 0.87
& 0.64
& 1 (all in scope by design)
& 1 \\
\bottomrule
\end{tabular}
}
\vspace{-5mm}
\end{table}


\textbf{Overall pass rate and analysis}.
\cref{fig:passatone} and \cref{fig:user_study} show that \textsc{VLP} consistently improves pass@1 under both LLM-simulated feedback and real user feedback. On BCB, \textsc{VLP} achieves the strongest results for both models, substantially outperforming other baselines. We attribute this advantage to our fine-grained mismatch detection through the linked documentation and prompts. Moreover, \textsc{VLP} exposes concrete implementation inconsistencies in the validation questions to users, and therefore eventually results in more actionable repair commands. Nevertheless, \textsc{VLP} still fails on 6.5\%--33.8\% of BCB tasks. Besides LLM inherent randomness, the remaining BCB failures are often caused by totally unstated operations that the prompts never mention. 
For example, in BCB-306, the prompt asks to remove all JavaScript files whose names contain  \texttt{jquery} and record the removed files in \texttt{jquery\_removal.log}; it never specifies that the implementation should use Python's logging module or call \texttt{logging.info}. Nevertheless, the BCB checker expects such a call, making this hidden requirement difficult for \textsc{VLP} to detect. Other failures arise when the correct feedback is not correctly applied by the repair LLM since it requires coordinated edits across scattered code snippets.

On QCE, \textsc{VLP} also improves the code generated by both models, especially for DeepSeek. However, it does not achieve full correctness because QCE tasks require substantial quantitative finance knowledge, such as reasoning about time-window alignment and look-ahead bias. \textsc{VLP} is not designed to detect inconsistencies beyond the LLM's knowledge and prompt. Furthermore, even when such inconsistencies are identified and the user provides the correct feedback, the repair model may still fail to implement the required changes, especially for DeepSeek. Overall, these results suggest that \textsc{VLP} is most effective when missing requirements or semantic deviations can be grounded in the program or prompt, while its remaining limitations stem primarily from the underlying model's domain knowledge and code repair capability.

\subsection{Simulated vs. Real User Consistency Analysis}
\label{subsec:user consistency}
Because simulator feedback affects the final pass@1, we examine whether it approximates real human feedback.
Note that TiCoder is excluded because users only need to accept or reject test cases without any NL interaction.
We use blind rubric-based adjudication, which is inspired by LLM-as-a-judge work that validates model-based outputs against human judgments~\cite{paper:llm_judge,paper:prometheus}.
For each task used in both LLM-simulated and real-user settings, we collect all question/comment-feedback pairs, anonymize them, and then ask three independent human adjudicators to score each response along two dimensions: (i) \emph{feedback correctness} measures whether the response is relevant and accurate with respect to the query or comment line (2: correct; 1: relevant but misleading; 0: irrelevant) and (ii) \emph{scope compliance} measures whether the response stays within the queried line (2: no leakage; 1: minor extra detail; 0: excessive cross-scope feedback).

We measure simulator-human consistency on each rubric dimension 
\(d \in \{\text{correctness}, \text{scope}\}\), corresponding to feedback correctness and scope compliance.
Let \(N\) be the number of questions/comments generated by GPT-5.4, and let each question/comment have one simulator response and \(J\) real-user responses.
For question/comment \(i\), we denote the simulator response as \(s_i\) and the \(j\)-th real-user response as \(h_{ij}\).
For each response, we aggregate scores from independent adjudicators by majority vote, using the median score to break ties; \(\hat r_d(x)\) denotes the resulting adjudicated score of response \(x\) on dimension \(d\).
We then compare the simulator score with the average human score:
\vspace{-7mm}

\[
\textsc{SimCons}_{d}
=
1-\frac{1}{N}\sum_{i=1}^{N}
\frac{\left|\hat r_d(s_i)-\frac{1}{J}\sum_{j=1}^{J}\hat r_d(h_{ij})\right|}{2}.
\]

\vspace{-2mm}
We also report average leave-one-out human-human consistency for reference.
The overall results in \cref{tab:simulator-validation} show that simulator feedback is reasonably aligned with human feedback.
Scope compliance is a bit less consistent for \textsc{VLP}, because fixing a bug may go beyond the scope of the documentation line directly associated with the question/comment.

\subsection{Quality of Validation Questions from VLP}
\label{subsec:precision}

In this section, we evaluate user validation burden and question usefulness. On BCB, \textsc{VLP} asks 8.79 questions per task on average with GPT-5.4 and 2.87 with DeepSeek V4 Flash, corresponding to 0.223 and 0.098 questions per line of code. On QCE, it asks 86.07 and 11.63 questions per task, with corresponding question-to-code-line ratios of 0.264 and 0.053. 
Among these questions, after review, 55.1\% and 64.1\% of cases on QCE and BCB require code modifications for DeepSeek, respectively, compared with 55.2\% and 66.3\% for GPT. Since GPT poses more questions, with a code-correcting question rate similar to DeepSeek, the simulated user plus GPT can cover more intent mismatches.
Compared with ClarifyGPT (0.015-0.164 questions per line) and PInG, \textsc{VLP} generally achieves a higher review-trigger rate than ClarifyGPT while eliciting substantially more user feedback, yet it requires far less review effort than PInG’s line-by-line comment review. Finally, question quality varies across LLMs used for detection, suggesting that the misalignment-detection prompt may require model-specific calibration.

After finishing a programming task, the participant answered 5 survey questions for the used method, with scores ranging from 1 (low) to 5 (high). The questions measure helpfulness for finding code issues, low labor intensity, understanding of the generated code, confidence in using the code beyond simple tests, and overall satisfaction. Higher scores indicate better user experience for all metrics. 
\cref{fig:user_study} shows that \textsc{VLP} achieves the highest scores in helpfulness for mismatch detection, code understanding, and confidence in using the code, while maintaining a lower labor-intensity than PInG and TiCoder. These results suggest that guided validation over code-derived documentation helps users validate LLM-generated code with a better balance between correctness and effort.

\begin{figure}[!t]
    \centering
    \includegraphics[width=\linewidth]{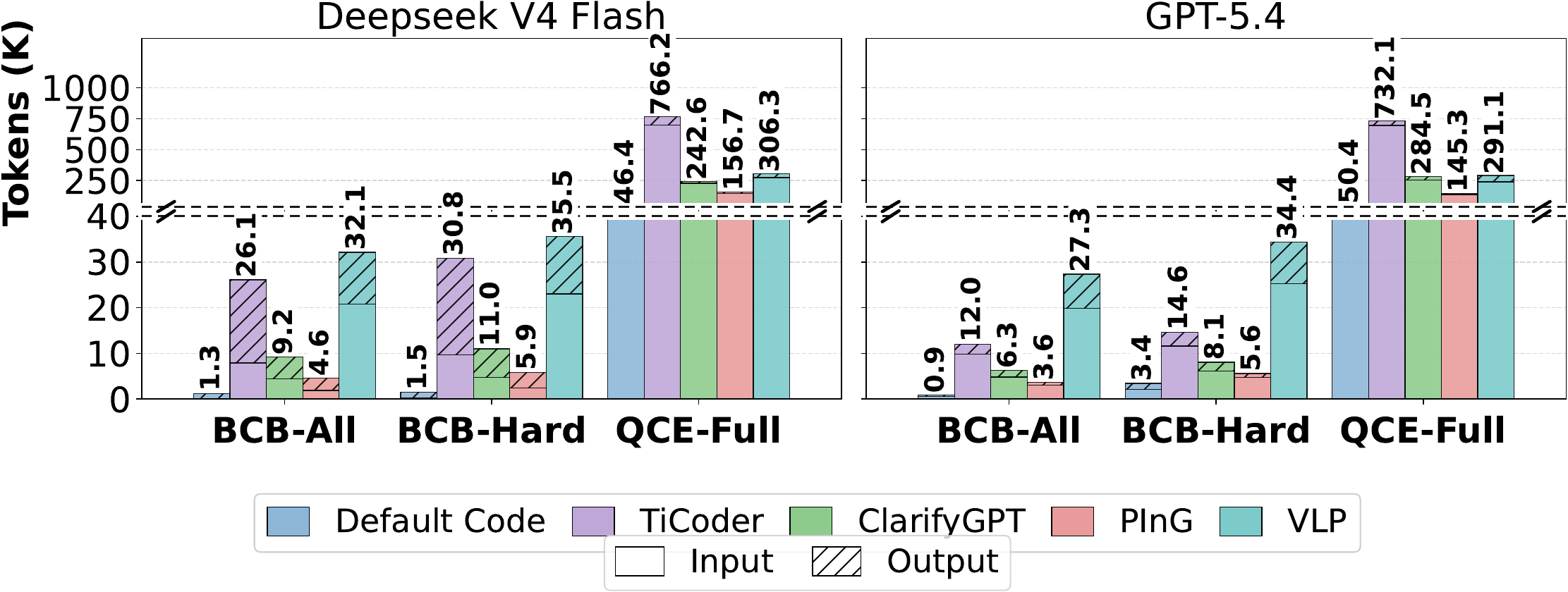}
    \vspace{-5mm}
    \caption{Average tokens spent on BCB and QCE.}
    \label{fig:tokencost}
    \vspace{-5mm}
\end{figure}

\subsection{Token and User Time Cost}
\label{subsec:cost-eff}

\cref{fig:tokencost} and \cref{fig:user_study} show the token cost and user pass@1 over time of each method, respectively. On BCB, \textsc{VLP} uses 27.3K tokens per task on average with GPT-5.4 and 32.1K with DeepSeek V4 Flash; on QCE, it uses 291.1K and 337.2K tokens, respectively. 
When we use DeepSeek and \textsc{VLP} to approach GPT's performance on complex tasks such as QCE, although 7.66$\times$ more tokens are used, the cost efficiency is still 2.33$\times$--6.99$\times$ higher.
In the real-user study, \textsc{VLP} slows users down more noticeably on short coding tasks, especially when using GPT, which asks more questions. However, for some more complex BCB and all QCE tasks used in real user study, where the code is longer, \textsc{VLP} almost lies on the Pareto frontier when considering both pass rate and user response time. It achieves higher correctness than faster methods and lower user time than methods with comparable correctness.

\subsection{Ablation Study}
\label{subsec:ablation}

We conduct an ablation study on BCB-Hard to measure the contribution of the API knowledge base and implementation-relevant TLR. With GPT-5.4, removing the API knowledge base decreases pass@1 from 91.9\% to 85.1\%, while removing TLR decreases pass@1 to 72.3\%. With DeepSeek V4 Flash, the two ablations both reduce pass@1 from 66.2\% to 58.8\%. These drops show that both components are important.

\subsection{Limitations and Threats to Validity}
First, although LLMs may have been trained on open-source dataset, data leakage is unlikely to substantially affect our results because \textsc{VLP} validates and repairs errors that remain in LLM-generated programs rather than reproducing benchmark solutions. Moreover, QuantCodeEval~\cite{paper:qcb}, released in May 2026, is unlikely to have appeared in model training data. Second, due to the cost of large-scale user studies, we combine a real-user study with an LLM-based user simulator. We improve simulation fidelity with clarifying questions and ground-truth test cases, and observe close agreement between simulated and real-user feedback (\cref{subsec:user consistency}). 
Finally, we do not evaluate the faithfulness of the documentation separately because most of it is enforced by a deterministic, syntax-directed translation. For API calls that require LLM-assisted translation, the LLM is provided with API documentation and constrained by our syntax, while the exact control flow, arguments, return targets, and data dependencies are preserved.


\section{Related Work}

\textbf{Validating/verifying LLM-generated code}.
Existing approaches to validating/verifying LLM-generated code mainly fall into two groups.
The first detects likely errors using probabilistic  methods, such as LLM-generated tests~\cite{paper:learn_to_verify,paper:rethinking_verification_for_code}, LLM judges, task-specific rules, risk prediction based on LLM internal states, calibration, or metamorphic testing~\cite{paper:verify_asr,paper:llm_check,paper:cali_and_correctness,paper:validate_llm_matamorphic}.
They can reveal failures, but cannot guarantee correctness.

The second group uses certificate-based verification, where generated code or specifications are checked by formal tools.
These methods rely on expert-written or LLM-generated properties~\cite{paper:property_based_testing,paper:learn_but_verify,paper:agent_guard,paper:sysspec,paper:fm-agent,paper:selfspec,website:speckit,paper:property_gpt}, verifier-guided refinement~\cite{paper:refine4llm}, or runtime checks with expert knowledge~\cite{paper:rvllm}.
Their key bottleneck is certificate quality: current LLMs often fail to produce properties that are both verifiable and complete enough to exclude incorrect behaviors~\cite{paper:evaluate_gpto4,paper:fv_with_mut_testing,paper:deepassert}.

\textbf{Human-in-the-loop for vibe coding}.
Human-in-the-loop techniques have been explored to make AI-assisted programming more controllable.
Prior work studies developer interactions with coding assistants~\cite{paper:grounded_copilot,paper:reading_between_lines,paper:ux}, elicits intent through intermediate feedback~\cite{paper:sketch_and_gen}, and updates generated code using runtime feedback or interactive model decisions~\cite{paper:live_programming_validation,paper:hilde,paper:hula}.
However, these systems mostly expose code snippets, tests, or runtime results to users.
\textsc{VLP} instead asks users to validate a human-readable NL representation of program semantics before verifying the code implementation.

Ambiguous or contradictory prompts can substantially degrade code correctness even when models still generate plausible code~\cite{paper:when_prompts_go_wrong}.
Existing work reduces underspecification mainly through pre-generation clarification or specification-oriented prompting~\cite{paper:clarifygpt,paper:ticoder,paper:clarigen,paper:whatpromptsdontsay}.
In contrast, \textsc{VLP} traces fine-grained links between the prompt and generated code, enabling iterative discovery of subtle mismatches.

\vspace{-1mm}
\section{Conclusion}
We present \textsc{VLP}, a human-in-the-loop framework that introduces an unambiguous literate documentation layer to bridge prompts and LLM-generated code. By enabling fine-grained misalignment detection, targeted user feedback, and automated verification over user-validated semantics, \textsc{VLP} improves code correctness with user validation effort, demonstrating that structured and user-friendly human-LLM collaboration is more effective than purely automated validation.


\bibliographystyle{IEEEtran}
\bibliography{ref}


\end{document}